\documentclass{aip-cp}[4apaper]

\usepackage[numbers]{natbib}
\usepackage{rotating}
\usepackage{graphicx}
\usepackage{lineno}
\begin{document}
%\linenumbers
% Title portion
\title{Measurement of the total cross section at 8 TeV and the inelastic cross section at 13 
TeV at the LHC with the ATLAS detector}

\author[aff1]{Hasko Stenzel \\ on behalf of the ATLAS Collaboration}
\eaddress{Hasko.Stenzel@cern.ch}

\affil[aff1]{II Physikalisches Institut, Justus-Liebig University Giessen, Heinrich-Buff Ring 16, D-35392 Giessen, Germany.}

\maketitle

\begin{abstract}
New measurements from ATLAS at the LHC are presented on the total cross section at $\sqrt{s}=8$ TeV from elastic scattering 
using the ALFA sub-detector and on the inelastic cross section at $\sqrt{s}=13$ TeV using the MBTS sub-detector.   
\end{abstract}

\section{Introduction}
The rise of the total proton-proton (pp) cross section with center-of-mass energy $\sqrt{s}$, predicted by Heisenberg
\cite{Heisenberg} and observed at the CERN Intersecting Storage Rings \cite{Pisa,ISR}, probes the non-perturbative regime
of quantum chromodynamics. Arguments based on unitarity, analyticity, and factorization imply
an upper bound on the high-energy behaviour of total hadronic cross sections that prevents them from
rising more rapidly than $\ln^2(s)$ \cite{froissart,martin,martin-inel}. The rise of the total and inelastic cross section 
in the energy range of the LHC is further confirmed by two new measurements from the ATLAS collaboration. At the highest centre-of-mass 
energy of 13 TeV a first measurement of the inelastic cross section is performed \cite{inelastic_13} and at 8 TeV the total cross 
section is determined \cite{elastic_8}.

\section{The inelastic cross section at $\sqrt{s}=13$ TeV}
\begin{figure}[ht!]\label{fig:nMBTS}
  \centerline{\includegraphics[width=8cm]{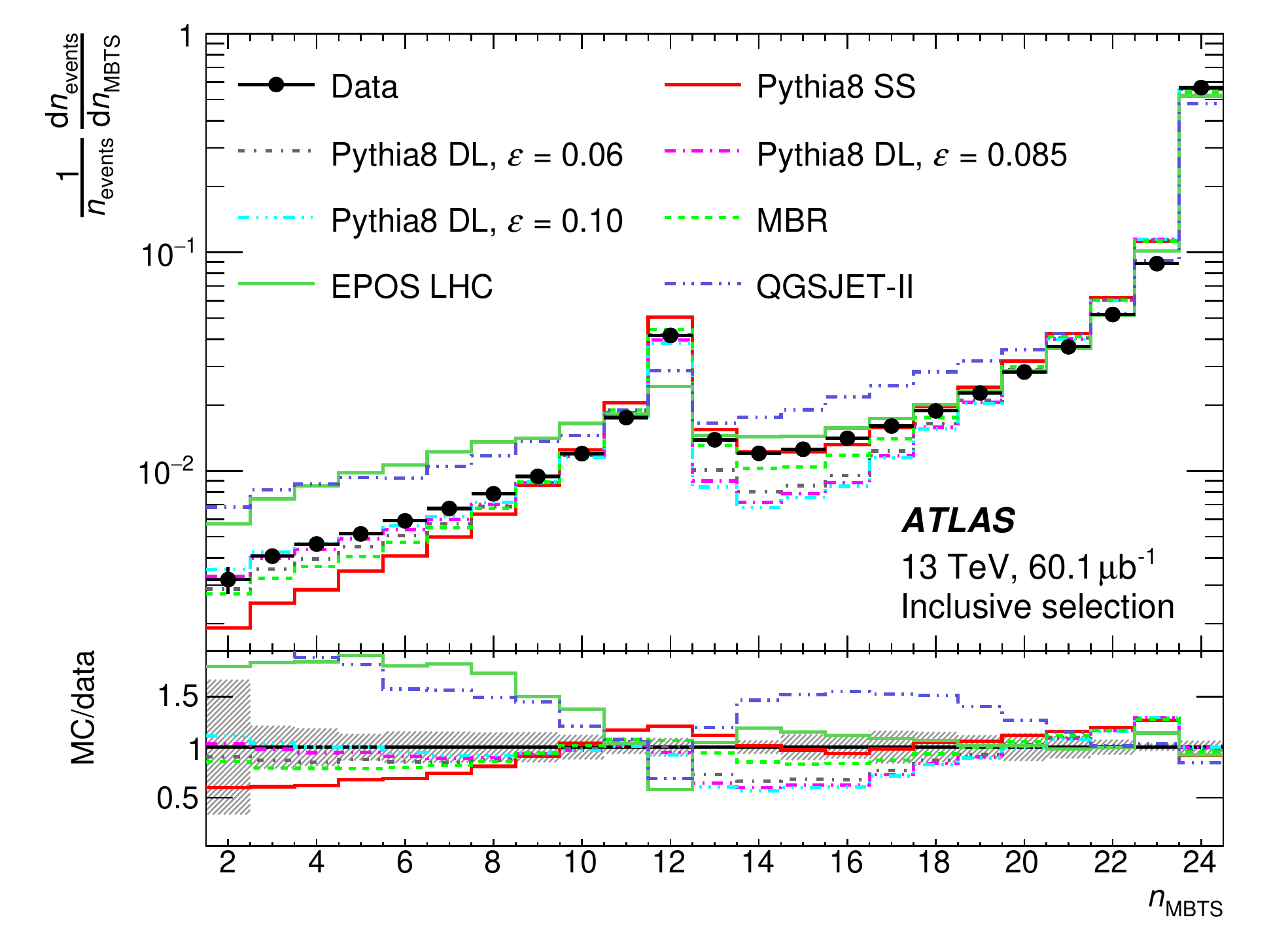}}
  \caption{The background-subtracted distribution of the number of MBTS counters above threshold in data and MC simulation 
  for the inclusive selection. The MC models are defined in Ref. \cite{inelastic_13} and their ratio to the data is also shown. The experimental uncertainty is shown as a 
  shaded band around the data points. The models shown here use the $f_{\rm D}$ value determined from the $R_{\rm SS}$ measurement. Figure from Ref.\cite{inelastic_13}.}
\end{figure}

The measurement of the inelastic yield is based on the counting of hits in the Minimum Bias Trigger Scintillator (MBTS), which is 
a plastic scintillator located in front of the end-cap calorimeters at a distance of 3.6 m from the interaction point in the forward region 
($2.07<|\eta|<3.86$). The counters are sensitive to diffractive dissociation processes with masses of the diffractive system 
$M_X>13$ GeV, or equivalently, $\xi=M_X^2 / s>10^{-6}$. Thus the inelastic cross section is primarily reported inside this 
fiducial volume corresponding to $\xi>10^{-6}$, and then extrapolated to full phase space with the help of models for the total 
inelastic cross section. Data were collected in a special run at low luminosity and low pile-up in which an integrated luminosity of 
about $60$ $\mu$b$^{-1}$ was accumulated. In order to select inelastic events nominally two hits in the MBTS are requested in any 
of the counters, additionally a single-diffraction enhanced sample is selected by requiring two counters with hits on the same side with a 
veto at the other side. The ratio of the number of events with a single-side tag and the number of inclusive events, $R_{\rm SS}$, is 
used to constrain in the simulation the relative contributions of single- and double-diffraction ($\sigma_{\rm SD}+\sigma_{\rm DD}$) to 
the inelastic cross section, $f_{\rm D}=(\sigma_{\rm SD}+\sigma_{\rm DD})/\sigma_{\rm inel}$, using a procedure developed in 
Ref. \cite{STDM-2010-11}. The resulting predictions for the MBTS hit distribution from different models, all tuned to reproduce 
the measured value $R_{\rm SS}=10.4 \pm 0.4 \%$, are shown in Fig. \ref{fig:nMBTS}.  
The fiducial cross section is determined by
\begin{equation}
\sigma_{\rm inel}^{\rm fid} \left( \xi > 10^{-6} \right) = \frac{ N - N_{\rm BG} } {\epsilon_{\rm trig} \times \cal{L} } \times \frac {1 - f_{\xi<10^{-6} } } { \epsilon_{\rm sel} },
\label{eq:main}
\end{equation}
\noindent where $N$ is the number of observed events passing the inclusive selection, $N_{\rm BG}$ is the number of background events, $\epsilon_{\rm trig}$ and $\epsilon_{\rm sel}$ 
are factors accounting for the trigger and event selection efficiencies, $1-f_{\xi<10^{-6}}$ accounts for the migration of events 
with $\xi < 10^{-6}$ into the fiducial region, and $\cal{L}$ is the integrated luminosity of the sample.
The measured fiducial cross section is determined to be 
$\sigma_{\rm inel}^{\rm fid} = 68.1\pm0.6\ \mbox{(exp.)}\ \pm1.3\ \mbox{(lum.)\ mb}.$
The extrapolation to $\sigma_{\rm inel}$ uses constraints from previous ATLAS measurements to minimize the model dependence of the component that falls outside 
the fiducial region. $\sigma_{\rm inel}$ can be written as

\begin{equation}\label{eqn:xsec}
\sigma_{\rm inel} = \sigma_{\rm inel}^{\rm fid} + \sigma^{\rm 7~TeV}(\xi<5\times10^{-6}) \times \frac{ \sigma^{\rm MC}( \xi<10^{-6} ) } {\sigma^{\rm 7~TeV, MC}(\xi<5\times10^{-6}) }.
\end{equation}

The term $\sigma^{\rm 7~TeV}(\xi<5\times10^{-6})=\sigma^{\rm 7~TeV}_{\rm inel}-\sigma^{\rm 7~TeV}(\xi>5\times10^{-6})=9.9\pm 2.4$~mb 
is the difference between $\sigma_{\rm inel}$ measured at 7~TeV using the ALFA detector~\cite{STDM-2013-10}, 
$\sigma_{\rm inel}^{\rm 7~TeV}$, and $\sigma_{\rm inel}$ measured at 7~TeV for $\xi>5\times 10^{-6}$ using the MBTS~\cite{STDM-2010-11}.
\begin{figure}[hb!]\label{fig:inelastic_running}
  \centerline{\includegraphics[width=8cm]{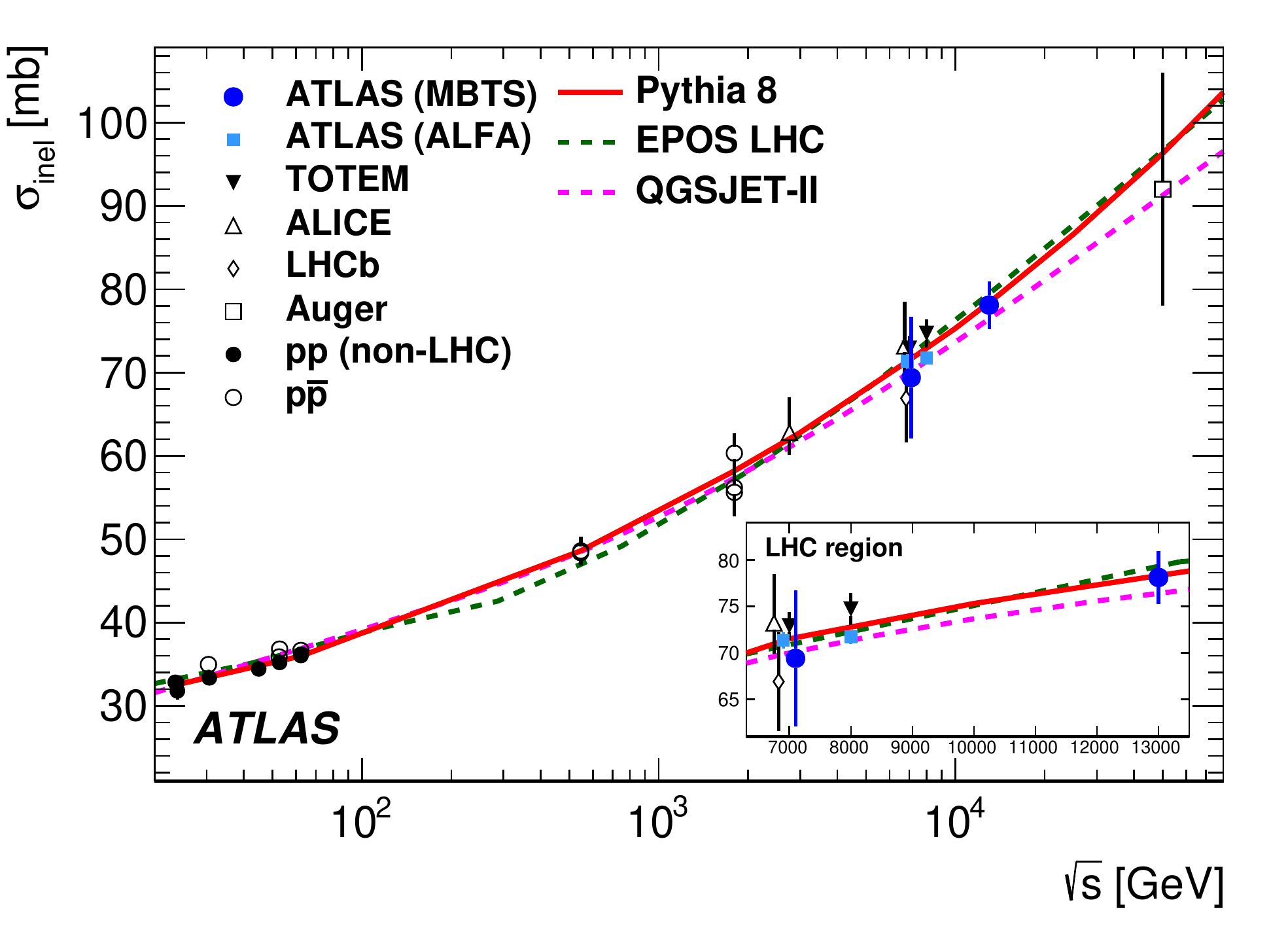}}
  \caption{The inelastic proton-proton cross section versus $\sqrt{s}$. Measurements from other hadron collider experiments and the Pierre Auger 
  experiment are also shown. The measurements are compared to the PYTHIA8, Epos LHC and QGSJet-II MC generator predictions. Figure from Ref.\cite{inelastic_13}.}
\end{figure}
This and other inelastic cross-section measurements are compared to several Monte Carlo models in Figure~\ref{fig:inelastic_running}. 
The measured value for $\sigma_{\rm inel}$ is
\begin{eqnarray*}
\sigma_{\rm inel} & = & 78.1 \pm 0.6~\mbox{(exp.)}\ \pm1.3~\mbox{(lum.)}\ \pm2.6~\mbox{(extr.)\ mb.} 
\end{eqnarray*}

% Head 2
\section{The total cross section at $\sqrt{s}=8$ TeV}
\begin{figure}[ht!]\label{fig:elastic_running}
  \centerline{\includegraphics[width=8cm]{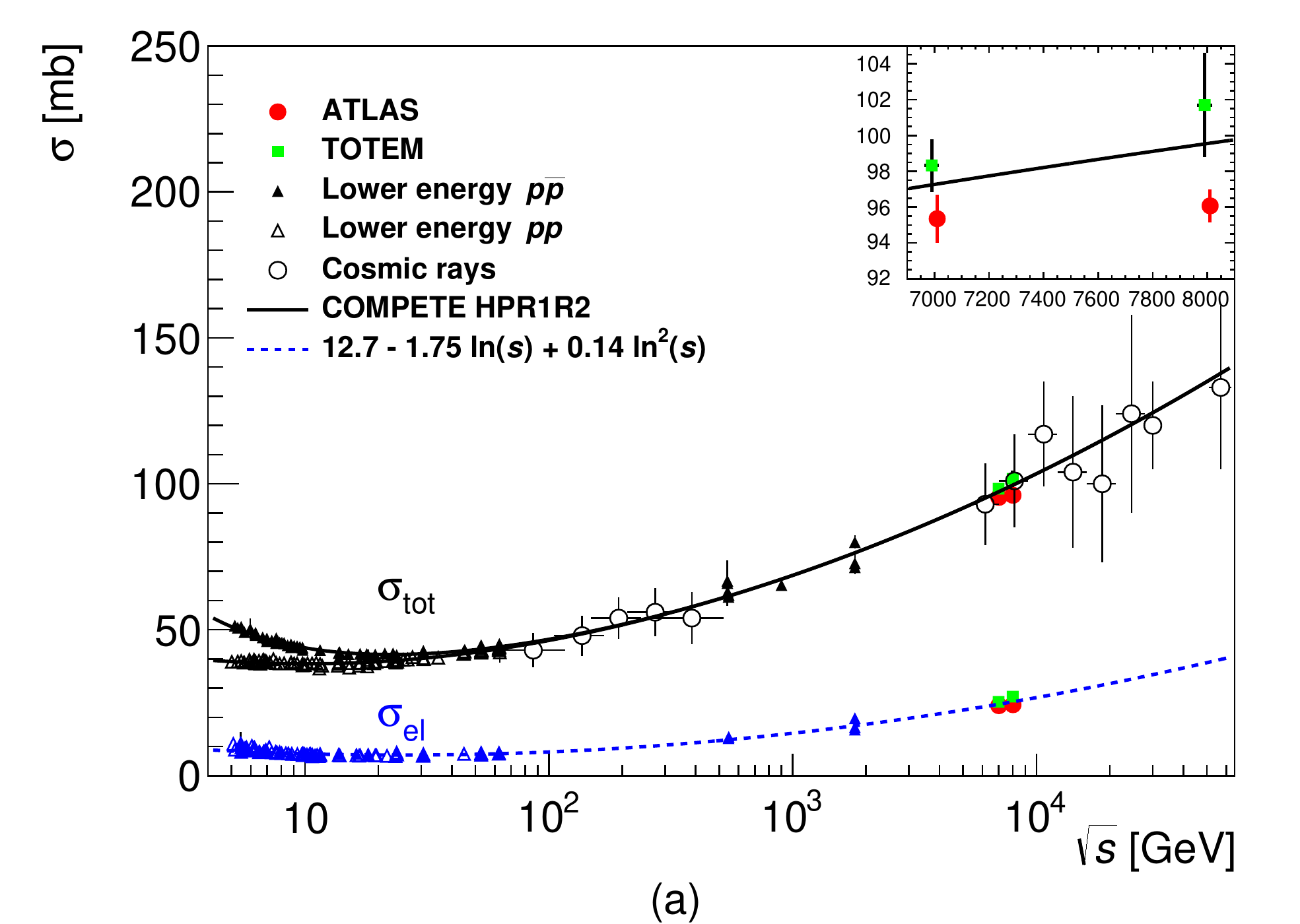}}
  \caption{Comparison of total and elastic cross-section measurements presented here with other published measurements and model 
  predictions as a function of the centre-of-mass energy. Figure from Ref.\cite{elastic_8}.}
\end{figure}

The total cross section is determined from elastic scattering data collected in a special run with $\beta^{\star}=90$ m using 
the ALFA Roman Pot sub-detector \cite{ALFA}. The optical theorem is exploited to extract the total cross section $\sigma_{\rm tot}$ 
from a fit to the differential elastic cross cross section according to
\begin{equation}\label{eq:totxs}
\sigma_{\rm tot}^{2} = \frac{16\pi(\hbar c)^2}{1+\rho^2} \left. \frac{\mathrm{d}\sigma_{\mathrm{el}}}{\mathrm{d}t}\right|_{t \rightarrow 0} \; ,  
\end{equation}
where $\rho$ represents a small correction arising from the ratio of the real to 
the imaginary part of the elastic-scattering amplitude in the forward direction.
Elastic events are selected with reconstructed tracks in four ALFA detectors with a trigger condition imposing a 
coincidence between the left and right arm of the set-up \cite{elastic_8}. Further cuts are imposed on the acollinearity of the 
events in order to reject background. The residual background contamination is determined from the data and subtracted. 
The resulting sample consisting about 3.8 M events. For each event the $t$-value is calculated according to 
$-t = \left(\theta^\star \times p \right)^2 \,$, where $p$ represents the nominal beam momentum and the scattering angle is 
calculated from the proton trajectories and beam optics parameters. The $t$-spectrum is unfolded for detector resolution and beam 
divergence effects and corrected for acceptance losses. The acceptance and unfolding corrections are calculated from a simulation 
tuned to account for the measured beam emittance and detector resolution \cite{elastic_8}. The rate of elastic events is further 
corrected for reconstruction inefficiencies occurring in events developing hadronic showers according to a data-driven procedure 
\cite{elastic_8}. Finally the integrated luminosity for this run is determined to be 
${L_\mathrm{{int}}} = 496.3 ~ \pm ~ 0.3 ~ (\mathrm{stat.}) ~ \pm ~ 7.3 ~ (\mathrm{syst.})~\mu$b$^{-1}$, where the systematic uncertainty 
of $1.5\%$ is dominated by the absolute luminosity scale calibration \cite{lumi}. 
The differential elastic cross section in a given bin $t_i$ is calculated from the following formula:
\begin{equation}\label{eq:cross-section}
\frac{\mathrm{d}\sigma_{\rm el}}{\mathrm{d}t_i} = \frac{1}{\Delta t_i}\times \frac{{\cal M}^{-1}[N_i - B_i]}{A_i  \times \epsilon^{\mathrm{reco}} \times \epsilon^{\mathrm{trig}} \times \epsilon^{\mathrm{DAQ}}  \times L_{\mathrm{int}} }\; \; , 
\end{equation}
where $\Delta t_i$ is the width of the bins in $t$, ${\cal M}^{-1}$ symbolizes the unfolding procedure applied to the 
background-subtracted number of events $N_i - B_i$, $A_i$ is the acceptance,  
$\epsilon^{\mathrm{reco}}$ is the event reconstruction efficiency, $\epsilon^{\mathrm{trig}}$ is the trigger efficiency, 
$\epsilon^{\mathrm{DAQ}}$ is the dead-time correction and $L_{\mathrm{int}}$ is the integrated luminosity. 
\begin{figure}[ht!]\label{fig:tfit}
  \centerline{\includegraphics[width=8cm]{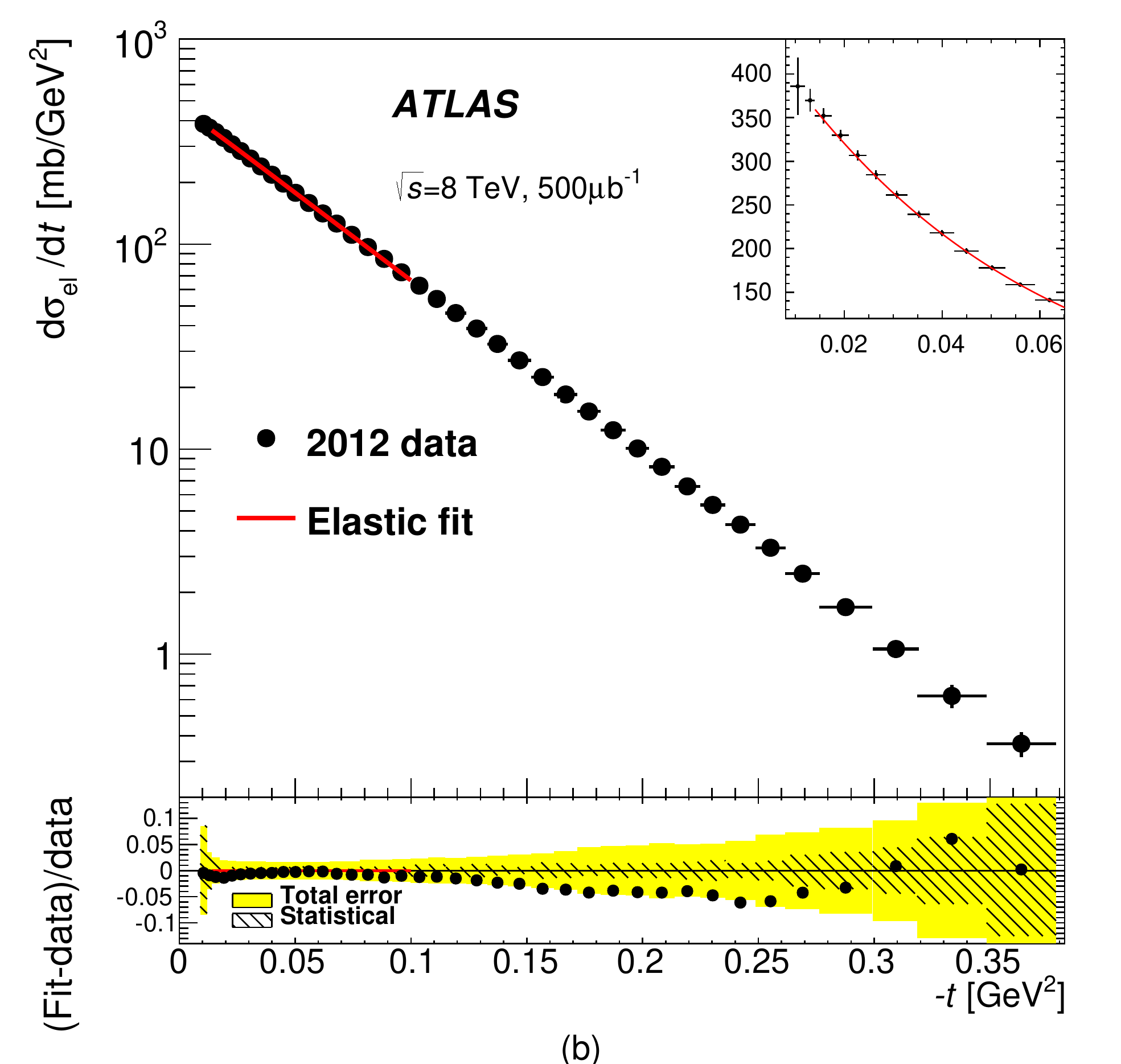}}
  \caption{The fit of the theoretical prediction to the
 differential elastic cross section with $\sigma_{\rm tot}$ and $B$ as free parameters.
In the lower plot the points represent the relative difference between fit and data, 
 the yellow area represents the total experimental uncertainty and the hatched area the statistical component. 
 The red line indicates the fit range; the fit result is extrapolated in the lower plot outside the fit range. 
 The upper right insert shows a zoom of the data and fit at small $t$. Figure from Ref.\cite{elastic_8}.}
\end{figure}
From a fit to the differential elastic cross section shown in Fig.~\ref{fig:tfit}, which included both statistical and experimental 
systematic uncertainties \cite{elastic_8}, the value of $\sigma_{\rm tot}$ and of the nuclear slope parameter $B$ are extracted. 
The fit yields for the  total cross section at $\sqrt{s}=8$ TeV 
\begin{eqnarray*}
\sigma_{\rm tot}(pp\rightarrow X) & = &  \mbox{96.07} \; \pm \mbox{0.18} \; (\mbox{stat.}) \pm \mbox{0.85} \; (\mbox{exp.})  \pm \mbox{0.31} \; (\mbox{extr.})  \; \mbox{mb} \; ,
\end{eqnarray*}
where the first error is statistical, the second accounts for all experimental systematic uncertainties and the last 
is related to uncertainties in the extrapolation $t\rightarrow 0$. 
In addition, the slope of the elastic differential cross section at small $t$ is determined to be  
$B = \mbox{19.74} \; \pm \mbox{0.05} \; (\mbox{stat.}) \pm \mbox{0.23} \; (\mbox{syst.}) \;  \mbox{GeV}^{-2}$. 
The dominant contributions to the systematic uncertainty arise from the luminosity and from the beam energy.  
The measurement of ATLAS is compared to measurements at lower energy and to a global  
fit in Figure~\ref{fig:elastic_running}.

\section{Conclusion}
The ATLAS Collaboration has performed new measurements of the total cross section at $\sqrt{s}=8$ TeV  
and of the inelastic cross section at $\sqrt{s}=13$ TeV. The main results are:
\begin{eqnarray*}
\sigma_{\rm tot}(\mbox{8 TeV}) & = &  \mbox{96.07} \; \pm \mbox{0.18} \; (\mbox{stat.}) \pm \mbox{0.85} \; (\mbox{exp.})  \pm \mbox{0.31} \; (\mbox{extr.})  \; \mbox{mb} \; , \\
\sigma_{\rm inel}(\mbox{13 TeV}) & = & 78.1 \pm 0.6~\mbox{(exp.)}\ \pm1.3~\mbox{(lum.)}\ \pm2.6~\mbox{(extr.)\ mb} \; .
\end{eqnarray*}

% References

\bibliographystyle{aipnum-cp}%
\bibliography{stenzel}%

\end{document}